\documentclass[pre,amsmath,showpacs]{revtex4-1}
\usepackage{graphicx} 
\usepackage{color}
\usepackage{bm}
\usepackage{epsfig}
\usepackage{latexsym}
\usepackage{pifont}
\usepackage{nameref,hyperref}
\begin{document}
\newcommand{\be}{\begin{equation}}
\newcommand{\ee}{\end{equation}}
\newcommand{\bea}{\begin{eqnarray}}
\newcommand{\eea}{\end{eqnarray}}

\def\Plus{\texttt{+}}

\title{Optimal methylation noise for best chemotactic performance of {\sl E. coli} }
\author{Subrata Dev and Sakuntala Chatterjee}
\affiliation{Department of Theoretical Sciences, S. N. Bose National Centre for Basic Sciences, Block - JD, Sector - III, Salt Lake, Kolkata 700106, India. }
\begin{abstract}
In response to a concentration gradient of chemo-attractant, {\sl E. coli} bacterium modulates the rotational bias of flagellar motors which control its run-and-tumble motion, to migrate towards regions of high chemo-attractant concentration. Presence of stochastic noise in the biochemical pathway of the cell has important consequence on the switching mechanism of motor bias, which in turn affects the runs and tumbles of the cell in a significant way. We model the intra-cellular reaction network in terms of coupled time-evolution of three stochastic variables, kinase activity, methylation level and CheY-P protein level, and study the effect of methylation noise on the chemotactic performance of the cell. In presence of a spatially varying nutrient concentration profile, a good chemotactic performance allows the cell to climb up the concentration gradient fast and localize in the nutrient-rich regions in the long time limit. Our simulations show that the best performance is obtained at an optimal noise strength. While it is expected that chemotaxis will be weaker for very large noise, it is counter-intuitive that the performance worsens even when noise level falls below a certain value. We explain this striking result by detailed analysis of CheY-P protein level statistics for different noise strengths. We show that when the CheY-P level falls below a certain (noise-dependent) threshold, the cell tends to move down the concentration gradient of the nutrient, which has a detrimental effect on its chemotactic response. This threshold value decreases as noise is increased, and this effect is responsible for noise-induced enhancement of chemotactic performance. In a harsh chemical environment, when the nutrient degrades with time, the amount of nutrient intercepted by the cell trajectory, is an effective performance criterion. In this case also, depending on the nutrient lifetime, we find an optimum noise strength when the performance is at its best. 
\end{abstract} \maketitle

\section{Introduction}
Behavior of a cell is controlled by the intracellular biochemical reactions in its signaling pathway. These reactions crucially depend on the expression levels of each protein involved and any fluctuations in the numbers of protein  molecules have important consequences on the cell performance \cite{jeschke,elowitz}. These fluctuations are also expected, since inside a single cell, the number of protein molecules which take part in the reactions,  is often small \cite{eldar} and can range between $10$ molecules per cell to $1000$ molecules per cell, depending on the type of the signaling protein \cite{b10}. How the variability in protein numbers affects the cell behavior is an important question to understand \cite{rao,tsimring,lan,book2014}.

A model system for studying cellular behavior is the chemotaxis pathway of {\sl E. coli} bacteria, which allows the cell to sense and respond to the changes in the nutrient concentration in its environment \cite{bergbook}. When the cell senses a concentration gradient, it moves up the gradient to reach a region of higher nutrient concentration \cite{adler1973}. The motion of an {\sl E. coli} is controlled by approximately ten flagellar motors, whose clockwise or counter-clockwise rotations generate run-and-tumble motion of the cell  \cite{berg3d,block}. When the motors rotate in the counter-clockwise (CCW) direction, the flagellar filaments associated with the motors form a helical bundle that smoothly propels the cell forward. This is known as a run. When one or two of these motors switch to a clockwise (CW) rotation, the flagellar filaments attached with them come out of the bundle and the motion becomes random with little net displacement. This is known as a tumble, at the end of which the cell begins a new run in a different direction. By modulating the rotational bias of the motors the cell can modulate its run and tumble durations \cite{bergcell}. When it moves up the concentration gradient of the  nutrient or a chemoattractant, the signaling network controls the motors in such a way that the runs are extended and the tumbles are suppressed \cite{bergcell}.

\begin{figure}[!h]
\includegraphics[scale=0.3]{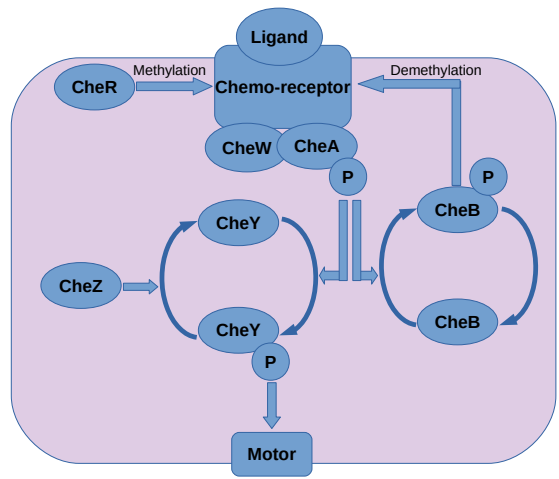}
\caption{Simplified chemotaxis pathway inside an {\sl E.coli} cell }
\label{fig:path}
\end{figure}

An external chemical signal is sensed by the chemo-receptors, which are proteins
bound to the membrane of the {\sl E. coli} cell \cite{soujikreview}. These receptors bind to the attractant molecules and in a bound state they suppress the phosphorylation of cytoplasmic protein CheA. The methylation level of the receptors is controlled by two proteins CheR and CheB: while CheR raises the level, the phosphorylated CheB-P causes demethylation to lower the level \cite{bren}. In a phosphorylated state, CheA transfers the phosphate group to CheY and CheB. When the phosphorylation of CheA is suppressed, CheB-P concentration also goes down and the demethylation process stops. The receptor then reaches a highly methylated state, which in turn raises the activity of CheA. This part of the network is responsible for adaptation. The other protein CheY which also receives phosphate group from CheA, controls the sensing mechanism of the network. CheY-P binds to flagellar motors and increases its CW bias, causing the cell to tumble \cite{eisenbach}. In the absence of phosphorylation, CheY-P concentration goes down which reduces CW bias and the cell swims smoothly. In Fig. \ref{fig:path} we present a simple diagram of the chemotaxis pathway.

In absence of any noise in the signaling pathway, the switching of rotational bias of the flagellar motors is expected to be a Poisson process and consequently, the duration of a particular run or tumble should follow an exponential distribution. Many early experiments and theoretical models which involve measurement over a bacterial population, indeed found exponential distribution \cite{bergcell,block}. However, in \cite{korobkova} the switching events of a single cell in a homogeneous nutrient environment were monitored in experiment and the residence time of the motors in the CCW biased state was found to follow a power law distribution. It was argued that the noise present in the signalling network of a single cell makes it possible to have large fluctuations in the CCW lifetimes and consequently, the cell can execute really long runs with significant probability \cite{tu2005,matthaus2009,matthaus2011,cluzel06,cluzel11}.

Interestingly, the large fluctuations present at the single cell level do not impair the chemotactic response \cite{emonet2008,park2010,sneddon2012,tu2013} or robust adaptation \cite{barkai,alon} observed at the population level \cite{emonet2016,manoj}. In a generic nutrient environment, cellular noise may prove beneficial, where large variability in a cellular population ensures that different cells behave in different ways and each type of behavior may be suitable for one particular type of environment \cite{frankel2014,shimizu15}. In \cite{he2016} the magnitude of the noise in the wild-type bacterial cell was estimated by comparing its behavioral difference with a mutant cell without signaling noise.

In \cite{flores} the effect of pathway noise on the chemotactic efficiency of a single cell was studied. The most important source of noise is the methylation-demethylation reactions in the network, since the time-scale of these reactions is order of magnitude larger than all other time-scales present in the pathway (see Sec \ref{sec:model}), the slow methylation fluctuations cannot be integrated out by the downstream processes in the network \cite{flores}. An optimal value of the network noise strength was reported for which the chemotactic performance of the cell is at its best \cite{flores}. In other words, starting from a very low signaling noise, as the noise is increased, the performance improves, before deteriorating again at large noise. To explain why performance improves with noise, it was argued that low activity states of the receptor mainly controls the chemotactic response. With increasing noise, these low activity states become more accessible and this was claimed to be the reason for noise-induced enhancement of chemotactic performance \cite{flores}.

In this paper, we further investigate the existence of optimal noise reported in 
\cite{flores}. In the presence of a nutrient concentration that varies in space but does not change with time, a good chemotactic performance may be described as the ability of the cell to climb up the concentration gradient quickly and localize in the nutrient-rich region in the long time limit. To characterize this performance, we measure two quantities, chemotactic drift velocity $V$ and localization $\langle C \rangle$. The chemotactic drift velocity is defined as the steady state average velocity with which the cell climbs up the chemical concentration gradient. Within the run-and-tumble motion of the cell, this drift velocity may be measured as the net uphill displacement of the cell during a run, divided by the average run duration. A large value of $V$ indicates a good chemotactic performance. The localization $\langle C \rangle$ is defined as the nutrient concentration, averaged over the steady state distribution of the cell position. This quantity takes a high value for a given concentration profile of the nutrient when in the long time limit most of the cells are present in the region which contains maximum nutrient. We find that both $V$ and $\langle C \rangle$ show peak as a function of signaling noise strength. 

To explain this observation, we perform a detailed analysis of CheY-P level fluctuations in presence of signaling noise. In particular, we monitor the net uphill displacement of a cell during a run as a function of CheY-P concentration measured at the beginning of the run. We find that for runs starting with low CheY-P levels, this displacement is negative, {\sl i.e.} the cell shows a tendency to move down the nutrient concentration gradient. This behavior is detrimental to chemotaxis and the threshold level of CheY-P below which this happens, decreases as noise increases. This explains noise induced enhancement of chemotactic performance. To the best of our knowledge, the existence of such a noise-dependent threshold of CheY-P level was not known earlier.

In the next section, we introduce the model in details. In Sec \ref{sec:br} we study the effect of methylation noise in a homogeneous nutrient concentration.
In Sec \ref{sec:space} we study the effect of signaling noise in a spatially varying nutrient concentration. Our conclusions are presented in Sec \ref{sec:con}. 

\section{Model description} \label{sec:model}
In \cite{tu2008,jiang}  the chemotactic pathway was modelled in terms of three dynamical variables, the activity $a(t)$ of the receptor complex, the methylation level $m(t)$ and the CheY-P level $y_P(t)$. In \cite{flores} this description was modified by incorporating methylation noise. In the present paper, we use the same model as in \cite{flores}.

The activity of a receptor complex is defined as the probability to find it in the active state. The free energy difference between the active and inactive state is denoted as $\epsilon (m, c(x))$, which is a function of the methylation level $m$ and the nutrient concentration $c(x)$ at the cell position $x$. Then within quasi-equilibrium approximation, activity can be written as 
\begin{equation}
a=\dfrac{1}{1+\exp[N \epsilon(m,c(x))]}, \label{eq:act}
\end{equation} 
where $N=6$ is the number of chemo-receptors participating in the signaling pathway. The free energy $\epsilon (m, c(x))$ can be written as a sum of contributions coming from $m$ and $c(x)$ as follows \cite{tu05,wingreen06}:
\be
\epsilon (m, c(x)) = \alpha(m_0-m) -\ln \left (\dfrac{1+c(x)/K_A}{1+c(x)/K_I} \right ). \label{eq:free}
\ee
As the cell position $x$ changes with time, the nutrient concentration $c(x)$ experienced by the cell also changes. In this work, we consider three different types of nutrient environment: ($i$) homogeneous concentration of nutrient, where $c(x)$ takes a constant value $c_0$ everywhere in the medium (see Sec. \ref{sec:br}), ($ii$) spatially varying nutrient concentration, considered in Sec. \ref{sec:space}, where $c(x)$ is assumed to be a linear or Gaussian function of the cell position $x$, ($iii$) nutrient concentration has an explicit time-dependence caused by diffusion and/or degradation of nutrient (see appendix \ref{sec:time}). In Eq. \ref{eq:free} $K_A =3 mM $ and  $K_I=18.2 \mu M$ \cite{jiang,flores} set the range of concentration that the cell is able to sense. The cell is insensitive to chemical concentration outside this range. The other parameter values are $\alpha =1.7$, $m_0=1$ \cite{jiang,flores}.

The methylation level of the receptor goes up under the action of the enzyme CheR (and goes down due to CheB-P). The concentration of CheR that is bound to the receptor fluctuates with time due to low abundance of the enzyme \cite{li04} and also due to binding-unbinding dynamics between free and bound state enzyme molecules \cite{li05,schul08,matthaus2009, matthaus2011}. This gives rise to fluctuations in the methylation level of the receptor. The resulting dynamics governing receptor methylation and demethylation is given by the stochastic equation \cite{flores}
\begin{equation}
\dfrac{dm}{dt}=k_R(1-a)-k_Ba+\eta(t). \label{eq:meth}
\end{equation}
Here, $k_R$ and $k_B$ denote the methylation and demethylation rate constants, respectively \cite{jiang} and $\eta (t)$ is the stochastic noise with properties $<\eta>=0$ and $<\eta(t)\eta(t^\prime)>= \lambda (k_R(1-\bar{a})+k_B\bar{a})\delta(t-t^\prime)$, where $\bar{a}=1/2$ is the average activity level in absence of any noise. The strength of the noise which determines the variance of methylation, depends on various biochemical rate constants and total concentration level of proteins like CheA, CheY and CheZ \cite{frankel2014}. While the rate parameters are generally expected to be constant for a given biochemical pathway, the total concentration level of different proteins can vary from cell to cell due to noisy gene expressions. Within our simple model, we vary the noise strength by varying the dimensionless parameter $\lambda$.  We consider only small values of $\lambda$ and within our range of variation, the fluctuation (measured as the standard deviation) in $m(t)$ remains significantly smaller than the average methylation value. The rate parameters $k_R$ and $k_B$ are significantly smaller than all other rates which characterize different reactions in the biochemical pathway. This makes the methylation fluctuation a slow process and hence the noise $\eta (t)$ cannot be integrated out. In our simulation, we have used $k_R=k_B=0.015 s^{-1}$ \cite{shimizu2010,flores}, which gives $<\eta(t)\eta(t^\prime)>= \lambda k_R \delta(t-t^\prime)$. Our main results remain unaffected even when $k_R$ and $k_B$ have small but different values.
 
Fluctuations in methylation level will also cause fluctuations in activity which in turn affects the phosphorylation of CheY proteins. In the phosphorylated state, CheY-P proteins bind to the flagellar motors and cause the cell to tumble. Denoting the fraction of phosphorylated CheY proteins as $y_P$, we can write \cite{flores}   
 \begin{equation}
\dfrac{dy_P}{dt}=k_Ya(1-y_P)-k_Zy_P,\label{eq:yeq}
\end{equation}   
where the phosphorylation and dephosphorylation rates of CheY molecules have the values $k_Y=1.7 s^{-1}$ and $k_Z=2 s^{-1}$ which are much higher than the
rates for methylation and demethylation \cite{tu2008,flores}. This is why no additive white noise in Eq. \ref{eq:yeq} has been included in the model, since it is expected that such noise would give rise to fluctuations much faster than that induced by methylation noise. The tumbling rate $\omega (y_P)$ is a sigmoidal function of $y_P$ \begin{equation}
\omega (y_P)= \Omega y_P^H \label{eq:omega}
\end{equation}
with $H=10$ and $\Omega = 282250s^{-1}$ \cite{cluzel2000,flores}. The value of $\Omega $ was estimated in \cite{flores} using the criterion that in an adapted state, the flagellar motors have a CW bias of $25 \%$. Although we use the same $\Omega $ value, in our simulation in one dimension, we consider instantaneous tumbling, {\sl i.e.} the cell does not spend a finite time in the tumbling state, but immediately after tumbling it starts running in a new direction. This is justified since the fraction of time spent in a CW bias state is negligible compared to that in CCW state. However, we have also checked that including a finite tumbling duration does not affect our conclusions.

In this paper, we consider motion of the bacterial cell in one and two dimensions. While one dimensional case is simpler to study and also relevant in view of recent experiments \cite{binz,tu17} where bacterial chemotaxis has been studied in narrow microfluidic channel inside which motion of the cell can be effectively considered to be one dimensional, we also verify that all our main results remain valid in two dimensions as well.

To perform simulations in one dimension, we consider a one dimensional box of length $L$, at the two ends of which there are reflecting boundary walls. In a time-step $dt$, the cell moves a distance $vdt$ where $v$ is the run speed. At the end of each step, the tumbling probability $\omega (y_P) dt$ is calculated and if a tumble does take place, the sign of $v$ is reversed with probability $q$. In each time-step the activity, methylation and CheY-P levels are updated according to Eqs. \ref{eq:act}, \ref{eq:meth} and \ref{eq:yeq}. Throughout we have used $L=1000 \mu m$, $v=10 \mu m /s$, $dt=0.01s$. To check for finite size effects we have also considered larger $L$ and smaller $dt$ values and found that our conclusions remain unaffected. 

For simulations in two dimensions, we consider an $L \times L$ box with reflecting boundaries in the $x$ and $y$ directions. The nutrient concentration gradient is assumed to be present only along the $x$ direction, but the cell moves on the $xy$ plane with velocity $\bf v$. The magnitude of $\bf v$ remains fixed but its direction changes randomly after each tumble. In this case we include finite tumble duration and rotational diffusion to make our model more realistic. The average tumble duration is taken to be $\tau_T = 0.1s$ and the rotational diffusivity $D_\theta = 0.062 \mu m^2/s$ \cite{berg3d,emonet2014,karmakar} allows gradual bending of the cell trajectory during a run. All other parameters remain same as in one dimensional case.

\section{Effect of signaling noise in a homogeneous nutrient environment} \label{sec:br}
We study the motion of a single cell in presence of a homogeneous nutrient concentration in the medium. Even in the absence of any concentration gradient of the nutrient, the effect of signaling noise is strongly felt.  A qualitative change in the run-length distribution is observed as the noise strength is varied. When the noise strength is zero, in a background of constant nutrient concentration, the activity level, methylation level and CheY-P level do not fluctuate and stay constant at their respective adapted values. The tumbling rate in Eq. \ref{eq:omega} then also takes a constant value and the bacterial motion consists of run and tumble modes with constant switching rates. The run-length distribution in that case in expected to be exponential. Over a length scale much larger than the average run-length, the motion of the cell can be described by a diffusion process.

\begin{figure}[!h]
\includegraphics[scale=1.4]{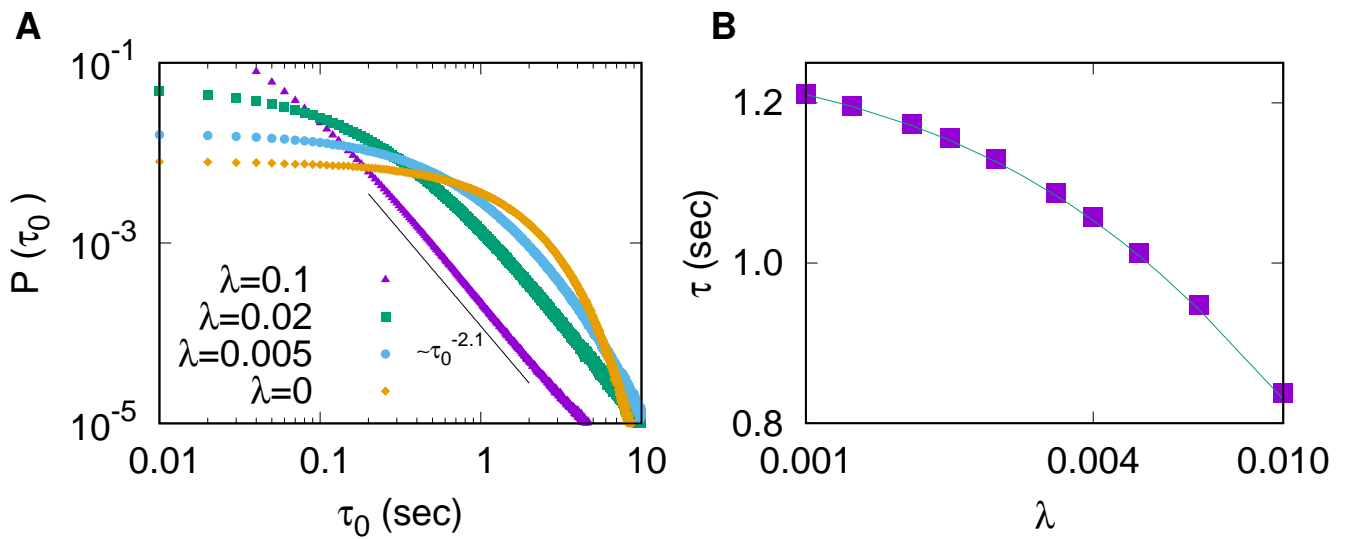}
\caption{{\bf Run duration statistics for different signaling noise.} {\bf (A)}: The probability $P(\tau_0)$ to observe a run duration $\tau_0$ for different noise strengths $\lambda$, in presence of a homogeneous nutrient concentration. For small $\lambda$, we find $P(\tau_0)$ has an exponential form, but for large $\lambda$ it is a power law with exponent $2.1 \pm 0.1$, close to experimental observation \cite{korobkova}. The thin line shows a power law function with power $2.1$. {\bf (B)}: The average run duration $\tau$ decreases with $\lambda$. The discrete points are for simulation and the continuous line shows analytical result. We find good agreement. Here, we have used a one-dimensional system and a homogeneous nutrient concentration with $c(x) = c_0 = 200 \mu M$. All other simulation parameters are as specified in Sec. \ref{sec:model}.} 
\label{power}
\end{figure}

On the other hand, when the noise strength is high, then methylation level in Eq.\ref{eq:meth} shows large fluctuations, which in turn induces fluctuations in the activity and in CheY-P level. The tumbling rate, which is a function of CheY-P concentration, also fluctuates with time. For large noise, the run-length distribution is known to decay like a power law with an exponent $\simeq 2.2$ \cite{korobkova} and the motion of the cell can be described by a L\'{e}vy walk \cite{matthaus2011}. A power law decay indicates the possibility of observing long runs in the system \cite{korobkova} and an exponent $2.2$ implies that although average run-length remains finite, the variance of the distribution diverges. The mean-squared displacement of the cell shows super-diffusive behavior in this limit. Within our model also, we verify the crossover of run-length distribution from an exponential to a power law with increasing noise strength in Fig. \ref{power}. Note that in our simulations we consider a finite system. This brings about an exponential cut-off in the tail of the run-length distribution which in turn restores diffusive behavior in the long time limit. We perform simulations in one dimension, but since the nutrient concentration is same everywhere in the medium, a two-dimensional system should also give the same results.

Although large noise increases the probability of very long runs, the average run-duration still decreases with noise, as shown in our data in Fig. \ref{power}B (discrete points). We will see later this has important consequence for the chemotactic response of the cell. 

\begin{figure}[!h]
\includegraphics[scale=0.7]{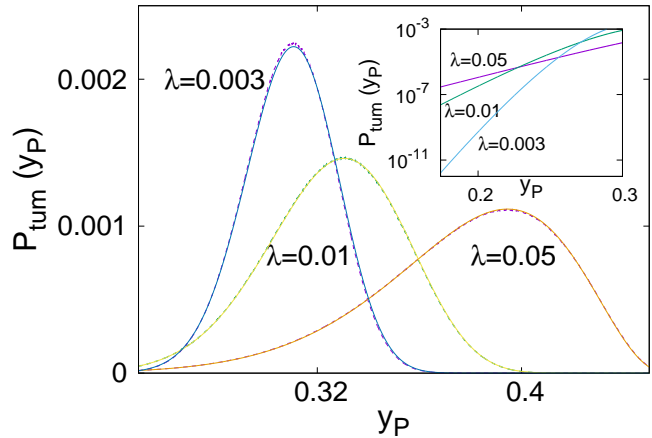}
\caption{{\bf The probability distribution $P_{tum} (y_P)$ for the fraction $y_P$ of phosphorylated CheY molecules for different noise strengths.}  With increasing noise strength, the peak  shifts rightward and the width increases. The dashed lines show (binned) simulation data and the continuous lines correspond to analytical calculation, which shows good agreement with simulation. Inset: The left tail region of the distribution on a zoomed scale. We find that, as noise increases, the tail becomes longer. The simulation parameters are as in Fig. 2. These data are for $c(x)=c_0$ and other simulation parameters are as in Fig. \ref{power}} .
\label{yp} 
\end{figure}

The crossover from exponential to power law shown in Fig. \ref{power} happens due to fluctuations present in the CheY-P level, which directly controls the motor bias \cite{tu2005}. To gain a deeper insight into this noise induced fluctuations, we measure the distribution of CheY-P concentration at the time of tumbling. Our simulation data in Fig. \ref{yp} shows that the CheY-P concentration follows a unimodal distribution. As noise increases, the distribution gets wider, as expected. Interestingly, the peak of the distribution shifts towards right with increasing noise, and the distribution develops a long tail for small values of CheY-P concentration. In the remaining part of this section, we outline steps to analytically calculate the distribution to explain these features. 

The fluctuation in the methylation level can be expressed in terms of the following stochastic differential equation \cite{gardiner}
\begin{equation}
dm=k_R \frac{c_0 e^{N \alpha (m_0-m)}-1}{c_0 e^{N \alpha (m_0-m)}+1}dt+\sqrt{{k_R\lambda}}dW(t),
\end{equation}
where $c_0$ is the uniform nutrient concentration in the medium and $dW(t)$ is a stochastic variable with uni-variate normal distribution. Using the expression of activity in Eq. \ref{eq:act} we can write  
\begin{equation}
da=k_RN\alpha a(1-a)(1-2a)(1+\frac{N\alpha\lambda}{2})dt+\sqrt{{k_R\lambda}}N\alpha a(1-a)dW(t) 
\label{eq:sde_a} \end{equation}
Let $F(a,t)$ denote the probability to find the cell with activity $a$ at time $t$. From Eq. \ref{eq:sde_a} we can construct the Fokker Planck equation for $F(a,t)$ which has the form
\begin{equation}
\frac{\partial F(a,t)}{\partial t}=-\frac{\partial}{\partial a}[k_RN\alpha a(1-a)(1-2a)(1+\frac{N\alpha\lambda}{2})F(a,t)]+\frac{k_R\lambda}{2}\frac{\partial^2}{\partial a^2}[N^2\alpha^2a^2(1-a)^2F(a,t)].
\end{equation}
In the steady state, the time-derivative on the left hand side vanishes and by making the transformation $ a=(1+u)/2 $ and $F(a) = [(1-u^2)/4]^{\kappa/2-1} G(a) $, with $\kappa=2/(N \alpha \lambda )$, the above equation gets reduced to associated Legendre equation
\begin{equation} 
(1-u^2)\frac{d^2G}{du^2}-2u\frac{dG}{du}+[\kappa(\kappa+1)-\frac{\kappa^2}{1-u^2}]G=0
\end{equation}
whose solution can be written as \cite{abra,wang_guo}
\begin{equation}
F(a)=\left [ \frac{1-(2a-1)^2}{4} \right ]^{\kappa/2-1}[A_1P^\kappa_\kappa(2a-1)+A_2 Q_\kappa ^\kappa(2a-1)] 
\end{equation}
where $P^\kappa_\kappa $ and $Q_\kappa^\kappa$ are  associated Legendre polynomial of first and second kind, respectively. The constants $A_1$ and $A_2$ can be determined from fitting with the numerical data. The distribution function for CheY-P follows from here, by noting that in steady state CheY-P concentration can be assumed to be equal to $\dfrac{a}{a+k_Z/k_Y}$. Here, we have used an assumption that the fluctuations present in activity $a$ are slow enough, such that for each level of activity, CheY-P level reaches a steady state. Then the probability that the fraction of phosphorylated CheY proteins has a given value $y_P$ is  
\begin{equation}
F_1(y_P)=F(a)\frac{da}{dy}=\frac{k_Y}{k_Z}(a+\frac{k_Z}{k_Y})^2 F(a)
\end{equation}
The probability distribution for CheY-P concentration at tumble can simply be calculated as $P_{tum}(y_P) = \omega(y_P)F_1(y_P)$. In Fig. \ref{yp} we compare our analytical results with simulations and find good agreement. As noise strength increases, the peak of the distribution shifts towards higher CheY-P level and the distribution also develops a long left-tail, as shown in the inset. We discuss below in Sec. \ref{sec:chey} that these facts play important role in understanding the chemotactic efficiency of the cell in the presence of nutrient concentration gradient.

The first moment of the distribution $P_{tum}(y_P)$  gives the average CheY-P level at tumble from which average tumbling rate can be calculated using Eq. \ref{eq:omega}, and the average run-duration can be estimated as the inverse of average tumbling rate. Note that in presence of a large signaling noise, the tumbling events are not Poissonian and this is only an approximate way to estimate the average run-duration. In Fig. \ref{power}B we compare our analytical result with simulations and find good agreement.

\section{Effect of signaling noise with spatially varying nutrient concentration}
\label{sec:space}
In the previous section, we have seen the tumbling rate and run duration get strongly affected due to noise, even when the cell moves in a homogeneous medium. It is expected therefore, that in presence of a concentration gradient of the nutrient, the chemotactic motion of the cell will also be seriously altered due to noise. In this section, we focus on what are the consequences of this on the chemotactic performance of the cell. We measure various different quantities which characterize different aspects of the chemotactic performance and study their properties for different strengths of the signaling noise in both one and two dimensions. We present in this section, our results on steady state chemotactic response characterized by localization and chemotactic drift velocity. In the appendix \ref{sec:fpt} we present our results on the first passage time of the cell that measures how quickly the cell manages to find the nutrient-rich region in the medium for the first time.  

\subsection{Steady state distribution of cell position} 
\label{sec:loc}
Let ${\mathcal P}_{\lambda}(x)$ be the steady state probability to find the cell in one dimension at position $x$, for a given noise strength $\lambda$. A good chemotactic performance implies strong localization of the cell in the nutrient-rich neighborhood. This means that ${\mathcal P}_\lambda (x)$ should be large whenever nutrient concentration $c(x)$ is large and ${\mathcal P}_\lambda(x)$ should take small value when $c(x)$ is small. A quantitative way to characterize this is to measure the average nutrient concentration experienced by the cell population in steady state, {\sl i.e.} $\langle C \rangle = \int_0^L dx c(x) {\mathcal P}_\lambda (x)$. Note that the integrand has a large value only when both $c(x)$ and ${\mathcal P}_\lambda (x)$ are large, indicating strong localization in favorable region. $\langle C \rangle - c_0$ indicates the difference between the average nutrient concentration captured by the cell and the background  nutrient concentration.

\begin{figure}[h]
\includegraphics[scale=0.5]{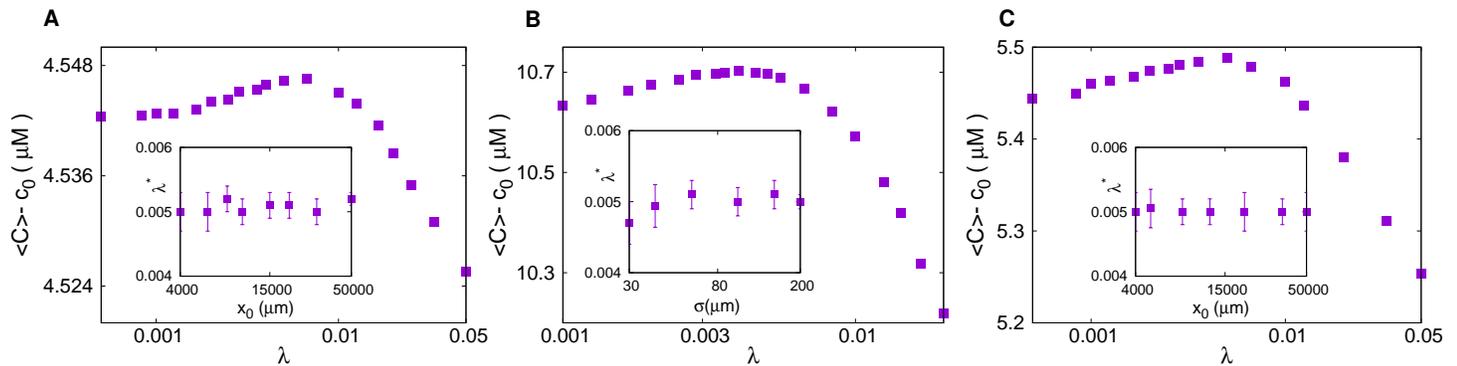}
\caption{{\bf Localization shows a peak as a function of noise strength.} {\bf (A)}: $ \langle C \rangle -c_0$ vs the noise strength $\lambda$ with $c(x)=c_0(1+x/x_0)$ with $x_0=20000 \mu m$. The optimum noise strength $\lambda^\ast \simeq 0.005$ in this case. Inset shows the plot for $\lambda^\ast$ vs $x_0$. We find no strong dependence of $\lambda^\ast$ on $x_0$. {\bf (B)}: The variation of $ \langle C \rangle - c_0$ with $\lambda$ for $c(x)=c_0(1+\frac{1}{\sqrt{2\pi\sigma^2}}\exp[-\frac{(x-\bar{x})^2}{2\sigma^2}])$, with $\sigma = 100 \mu m$, $\overline{x} = 500 \mu m$. This case also shows similar value for $\lambda^\ast$. The inset shows the plot of $\lambda^\ast$ vs $\sigma$. {\bf (C).} $ \langle C \rangle -c_0$ vs the noise strength $\lambda$ in two dimension with $c(x)=c_0(1+x/x_0)$ with $x_0=20000 \mu m$. The localization shows a peak at the same value as in {\bf(A)} and {\bf(B)}. The inset shows the variation of $\lambda^*$ with the $x_0$. For two dimension case also we find no strong dependence of the optimum noise strength on the gradient present in the system. In all panels we have used $c_0 = 200 \mu M$ and other simulation parameters are as specified in Sec \ref{sec:model}.} 
\label{loc} \end{figure}

We find that $\langle C \rangle -c_0$ shows a non-monotonic variation with noise strength $\lambda$: while for very small and very large $\lambda$ values $\langle C \rangle$ is low, for intermediate noise level, $\langle C \rangle$ reaches a peak. This means that there is an optimum level of the signaling noise when the chemotactic performance, as measured by $\langle C \rangle$, is at its best. In Fig. \ref{loc}A, we present the data for the linear concentration profile and find that the best chemotaxis is observed for $\lambda = \lambda^\ast \simeq 0.005$.  The value of $\lambda^\ast $ does not seem to depend strongly on the concentration gradient (see inset of Fig. \ref{loc}A). In Fig. \ref{loc}B we show the data for a Gaussian form of $c(x)$ which also shows a comparable value of $ \lambda^\ast$. In Fig. \ref{loc}C we present our results for the two dimensional case with a linear concentration profile $c(x)$ and find similar behavior. 

Although localization does reach a peak at $\lambda^\ast$, the peak is not so pronounced. For a linear $c(x)$, our choice of large $x_0$ ensures a weak gradient and this yields a linear ${\mathcal P}_{\lambda}(x)$. The slope of this distribution  can be used as another characteristic to measure the chemotactic performance. As expected, this slope in one dimension also shows a peak at the same $\lambda^\ast$ value (data in appendix \ref{appendixB} Fig. \ref{S1_Fig}) and this peak is much more pronounced. 
\begin{figure}[!h]
\includegraphics[scale=0.5]{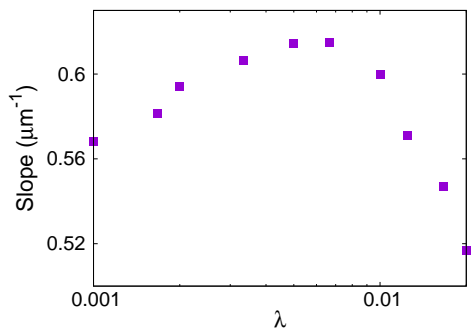}
\caption{ For a linear nutrient concentration with weak gradient, $P_\lambda (x)$ is also linear and its slope shows a peak as a function of noise strength. The optimum noise strength $\lambda^\ast$ is close to  0.005. Here, we have used $c(x)=c_0 (1+x/x_0)$ with $x_0 = 20000 \mu m $, and we have scaled the slope by a factor of $10^8$. All other simulation parameters are as in Fig. 4(a).}
\label{S1_Fig} \end{figure}

\subsection{Chemotactic drift velocity in steady state}
\label{sec:drift}
The signaling network inside the cell is such that the runs in the direction of increasing concentration gradient of the nutrient are extended and those in the opposite direction are shortened. This gives rise to an overall drift motion up the concentration gradient \cite{degennes}. Even in the absence of any methylation noise, the inherent stochasticity in the run-tumble motion of the cell gives rise to an effective diffusion in the long time limit which tends to homogenize the cell population. It is the drift motion which helps the system sustain the spatial variation in the steady state population density ${\mathcal P}_\lambda (x)$  \cite{sc2011,kafri2008}. In the presence of methylation noise, the cell trajectories may show super-diffusive behavior, which would again flatten out ${\mathcal P}_\lambda (x)$, had there been no drift motion. Therefore, chemotactic drift is crucial for chemotactic response. A large value of the chemotactic drift velocity means that the cell can quickly climb up the concentration gradient. It is certainly an important criterion for chemotactic performance.

We measure the chemotactic drift velocity of the cell in one dimension in presence of a linear concentration profile of the nutrient. Before presenting our simulation data, we include a brief discussion on how we measure the drift velocity from the run-and-tumble trajectory of the cell. Note first that the existence of a non-zero drift velocity means that the average run duration in the rightward direction (up the gradient) is different from that in the leftward direction (down the gradient). We measure this difference at an arbitrary position $x$ where the cell tumbles and a new run begins,  and finally average over all $x$ values. Let $N_R(x)$ and $N_L(x)$ be the total number of rightward and leftward runs starting at $x$, within an observation time window $t_{obs}$. Let $d_R(x)$ and $d_L(x)$ be the total durations of these rightward and leftward runs. The average run duration (in either direction) starting at $x$ is then given by $\tau(x)=\dfrac{d_R(x)+d_L(x)}{N_R(x)+N_L(x)}$. Note that $\tau(x)$ is in general different from $[\tau_R(x)+\tau_L(x)]/2$, where $\tau_R(x)$ is the average duration of a rightward run starting at $x$ and is equal to $d_R(x)/N_R(x)$. Similarly, $\tau_L(x)=d_L(x)/N_L(x)$. The difference stems from the fact that $N_R(x)$ and $N_L(x)$ are not equal in general.

\begin{figure}[!h]
\includegraphics[scale=0.5]{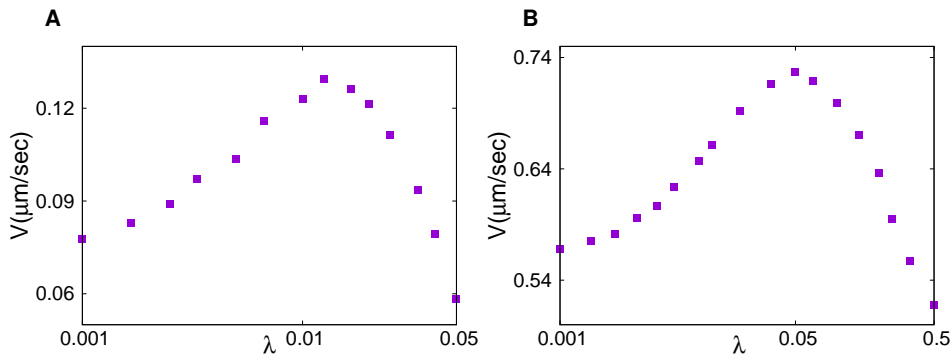}
\caption{{\bf Chemotactic drift velocity shows a peak as a function of the noise strength.} Left and right panels show the data for one dimensional and two dimensional systems, respectively. In both cases, the optimal noise strength $\lambda_o$ is found to be higher than that for localization data in Fig. \ref{loc}. We have used $c(x)=c_0(1+x/x_0)$ here and all simulation parameters are as in Fig. \ref{loc}.}
\label{drift} \end{figure}

\begin{figure}[!h]
\includegraphics[scale=0.5]{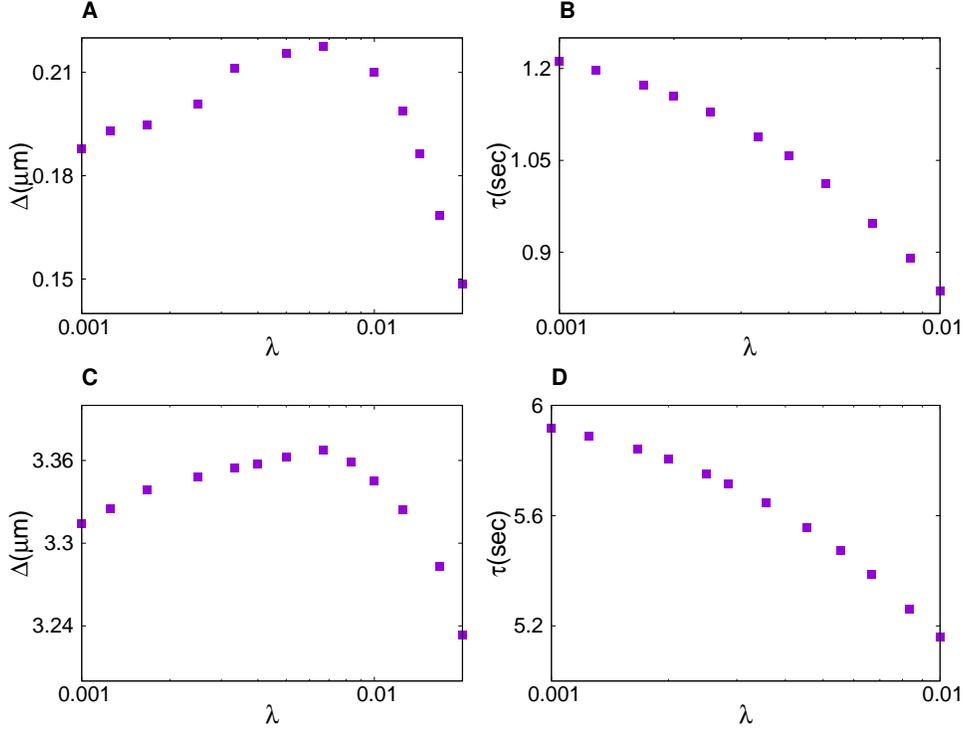}
\caption{{\bf The net displacement $\Delta$ in a run and the average run-duration $\tau$ as a function of noise strength.} $\Delta$ shows a peak at an optimum noise value close to $\lambda^\ast$, and $\tau$ decreases monotonically with noise. The upper panel shows data for one dimension and the lower panel shows data for two dimensions. All the simulation parameters for one and two dimension are same as in Fig. \ref{drift}.} 
\label{fig:tau} \end{figure}

The probability that a run starts from the position $x$ is $Q_{tum}(x)={\cal N}^{-1}[N_R(x)+N_L(x)]$ with the normalization constant ${\cal N}=\int dx' [N_R(x')+N_L(x')]$.  The average displacement in a run can then be calculated as $\Delta = \int dx Q_{tum}(x) v\dfrac{d_R(x)-d_L(x)}{N_R(x)+N_L(x)}$. The chemotactic drift velocity is obtained on dividing the average displacement in a run by the average run duration $\tau=\int dx \tau(x) Q_{tum}(x)$. Thus the final expression for chemotactic drift velocity \cite{emonet2014} is 
\be
V=\frac{\Delta}{\tau}=\frac{v\int dx [d_R(x)-d_L(x)] }{\int dx' [d_R(x')+d_L(x')]}
\ee
In Fig. \ref{drift} we plot $V$ for different noise strengths $\lambda$ and find that $V$ shows a peak as a function of $\lambda$. The position of the peak does not match exactly with what we observed for localization $\langle C \rangle$. The chemotactic drift velocity reaches a peak value for an optimum noise strength $\lambda_o \simeq 0.01$, somewhat higher than the optimum noise strength $\lambda^\ast$ for $\langle C \rangle$. To explain this difference, we separately plot $\tau$ and $\Delta$ as a function of noise in Fig. \ref{fig:tau}. We find that $\tau$ decreases monotonically with noise, as in a homogeneous nutrient environment (also see Fig. \ref{power}B). However, $\Delta$ shows a peak at a noise value, which again matches with $\lambda^\ast$. Although the non-monotonic variation of $V$ with noise arises due to that of $\Delta$, we can also see why the peak of $V$ occurs at a higher noise value. Using $V=\dfrac{\Delta}{\tau}$, at the peak position $\lambda_o$ one must satisfy the condition that $\tau \Delta ' -\Delta \tau '=0$, where the primes denote derivative with respect to $\lambda$. Since $\tau'<0$ for all $\lambda$, it immediately follows that $\Delta' <0$ at $\lambda=\lambda_o$. In other words, $\Delta$ decreases with noise at $\lambda=\lambda_o$, which means it has reached its peak at a smaller $\lambda$ value. We verify all these results in two dimensions as well. Here, the nutrient concentration gradient is applied along $x$-direction and hence the chemotactic drift is also present only in  $x$-direction. The motion of the cell along $y$-direction is expected to be purely diffusive in this case. We present our results for $V$, $\Delta$ and $\tau$ as a function of $\lambda$ in Figs. \ref{drift}B, \ref{fig:tau}C and  \ref{fig:tau}D, respectively, for the two dimensional case.

The fact that the peak position of $\Delta$ matches with that of  $\langle C \rangle$  indicates that there is indeed a unique noise strength at which the chemotactic performance of the cell is at its best. Moreover, this also shows that the chemotactic response is drift mediated and to understand the origin of optimality, we need to examine the noise dependence of $\Delta$ in detail. We discuss this in the following subsection. For the sake of simplicity, we limit our discussion to the one dimensional case here, but our arguments can be generalized to the two dimensional case as well.

\subsection{Explanation of optimal noise strength}
\label{sec:chey}	
\begin{figure}[!h]
\includegraphics[scale=1.0]{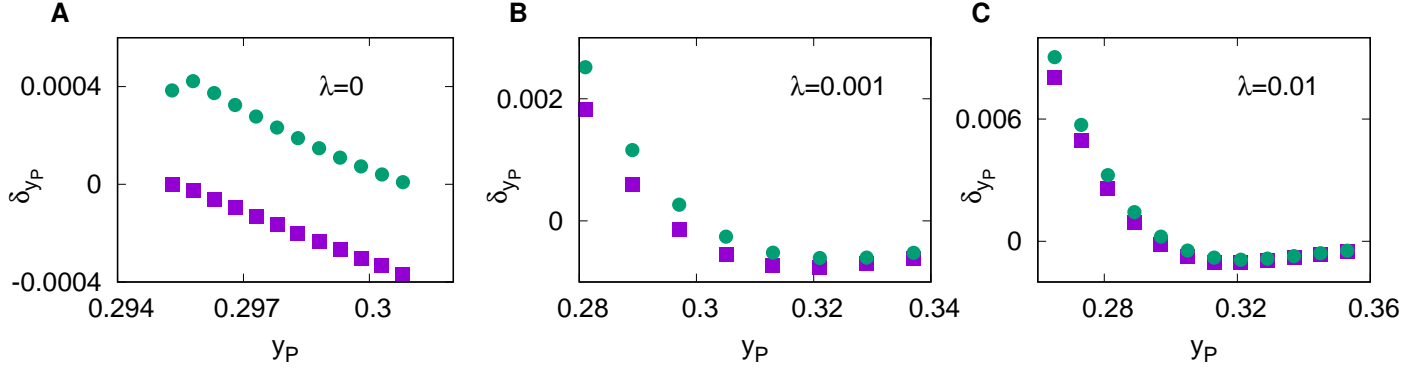}
\caption{{\bf Average change in CheY-P level during a run, as a function of the initial CheY-P level, at the start of the run.} $y_P$ denotes the fraction of phosphorylated CheY molecules and $\delta_{y_P}$ denotes its average change. The circles (squares) show the data for leftward (rightward) runs. {\bf(A)}: In absence of methylation noise, average CheY-P level always goes down (up) during a rightward (leftward) run. {\bf(B)}: For intermediate noise strength, change in CheY-P becomes positive (negative) during a rightward (leftward) run for small (large) values of CheY-P concentration. {\bf(C)}: As noise increases, the difference between the two curves become smaller. These data are for the one dimensional system and the simulation parameters are same as in Fig. \ref{drift}A.} \label{fig:chy} \end{figure}

First let us consider the case of very low methylation noise. In this limit, the only source of fluctuations in activity $a(t)$, methylation $m(t)$ or CheY-P level $y_P(t)$, is the change in the position of the cell. As the cell moves rightward (leftward), the concentration $c(x)$ of the chemo-attractant increases (decreases), and as follows from Eqs. \ref{eq:act} and \ref{eq:free}, the activity decreases (increases). This in turn, causes CheY-P level $y_P(t)$ to go down (up) in a rightward (leftward) run. In our simulations, we record the value of the CheY-P concentration at the beginning of a run and measure the average change in $y_P(t)$ at the end of that run. In Fig. \ref{fig:chy}A, the lower (upper) curve shows the data for a rightward (leftward) run for noise strength $\lambda=0$. This plot shows that rightward runs bring down the CheY-P level and leftward runs push the level up \cite{emonet2014}. 

However, as $\lambda$ increases, the change in $a(t)$ or $y_P(t)$ is not solely controlled by the change in cell positions, but also by the methylation level fluctuations. Our data in \ref{fig:chy}B show that for rightward runs (purple squares), change in $y_P(t)$ is negative for large $y_P$, but positive for small $y_P$. Similarly, for leftward runs (green circles), change in $y_P(t)$ is positive for small $y_P$, but negative for large $y_P$. This happens due to the feedback mechanism: when $y_P$ becomes too low (too high), the methylation level increases (decreases) to bring the $y_P$ level up (down). For large noise, when the range of variation of $y_P$ is larger, the feedback effect is more prominent.

\begin{figure}[!h]
\includegraphics[scale=1.2]{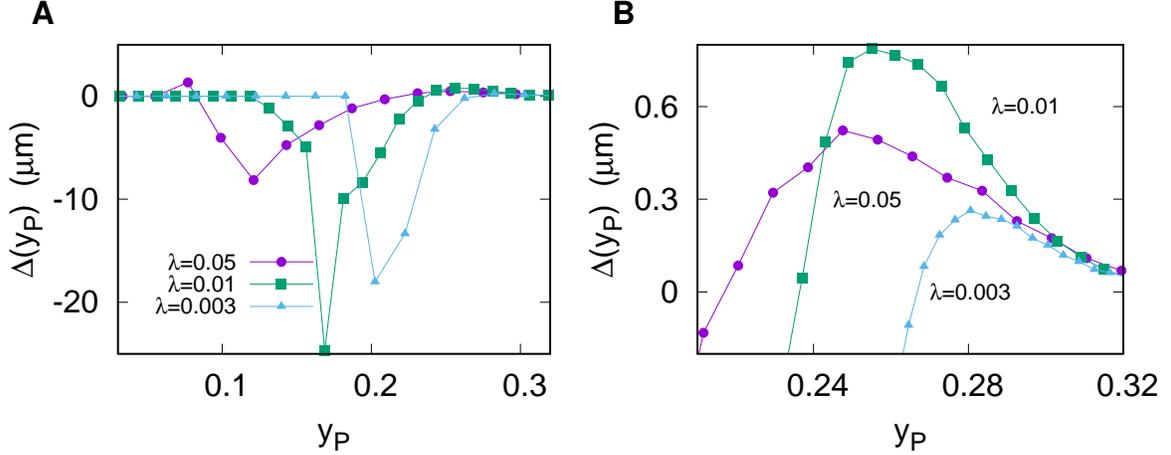}
\caption{{\bf Net displacement during a run for different noise strengths.} $y_P$ denotes the fraction of phosphorylated CheY molecules, at the start of the run, and $\Delta(y_P)$ denotes the average displacement in that run. The left panel shows the complete range of $y_P$ values, while the right panel zooms onto the large $y_P$ values. {\bf(A)}: $\Delta(y_P)$ vanishes for very small or very large $y_P$ values and attains a large negative peak and small positive peak for intermediate $y_P$ values. The position and height of the peaks depend strongly on noise. {\bf(B)}: The positive peak of $\Delta(y_P)$  shown on a zoomed scale. While the position of the peak shifts towards left as noise increases, the height of the peak clearly shows non-monotonic behavior with noise, the highest peak being observed close to the optimal noise $\lambda ^\ast$. These data are for one dimension and all simulation parameters are same as in Fig. \ref{drift}A.} 
\label{fig:yp}
\end{figure}

Let us now turn our attention to the quantity $\Delta$, the net displacement of the cell position in a run. In the previous subsection, we had measured $\Delta$ in terms of position dependent quantities like $d_R(x), N_R(x)$, etc. and then averaged over all $x$. Alternatively, one can measure these quantities as a function of $y_P$ and do a weighted average with $P_{tum}(y_P)$ over different $y_P$ values. This approach may be particularly instructive since CheY-P directly controls the tumbling rate of the cell. For this purpose, we define $\Delta (y_P)= \dfrac{d_R(y_P)-d_L(y_P)}{N_R(y_P)+N_L(y_P)}$, where $N_R(y_P)$ ($N_L(y_P)$) denotes number of rightward (leftward) runs starting with CheY-P concentration value $y_P$ and $d_R(y_P)$ ($d_L(y_P)$) denotes the total rightward (leftward) displacement of the cell position in these runs. Clearly, weighted average of $\Delta (y_P)$ with the distribution $P_{tum}(y_P)$ over different $y_P$ values gives back the same $\Delta$ as shown in Fig. \ref{fig:tau}A. We plot $\Delta (y_P)$ as a function of $y_P$ for different noise strengths in Fig. \ref{fig:yp}. 

\begin{figure}[!h]
\includegraphics[scale=1.3]{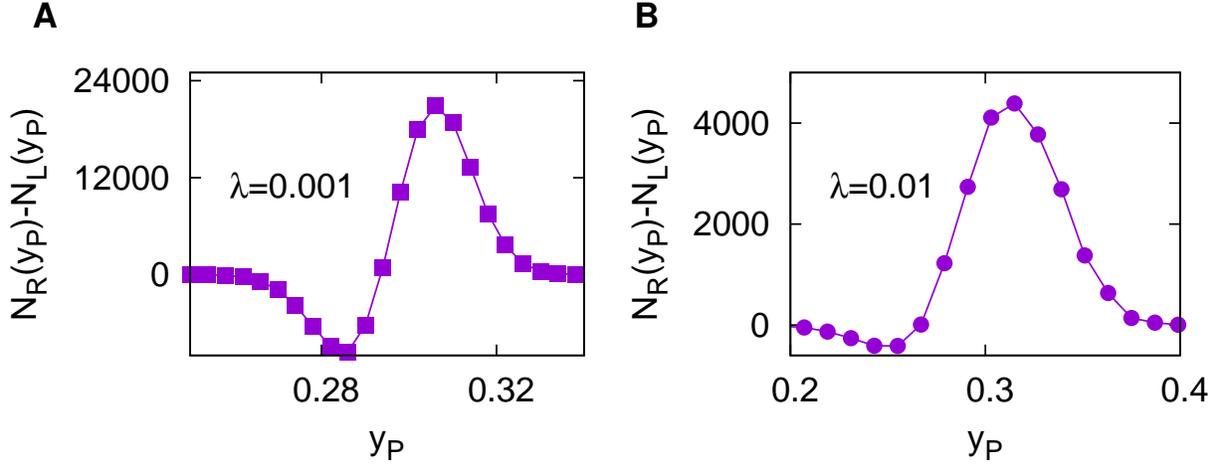}
\caption{{\bf Variation of $N_R(y_P)-N_L(y_P)$ against $y_P$ for different noise strengths.} {\bf (A):} As $y_P$ increases, $N_R(y_P)-N_L(y_P)$ starts from zero, reaches a negative peak, followed by a positive peak, and again becomes zero for very large $y_P$ values. The negative values observed at small $y_P$ show that in this range, the cell motion is biased towards decreasing nutrient concentration. {\bf (B):} Similar behavior is observed at large $\lambda$, but the point of zero-crossing shifts towards smaller $y_P$ values. All simulation parameters are same as in Fig. \ref{drift}A.}
\label{S2_Fig} \end{figure}

For low noise, our data in Fig. \ref{fig:yp} show that $\Delta (y_P)$ is negative for small $y_P$, increases to a positive value as $y_P$ increases, then reaches a peak and finally decays to zero for large $y_P$. A negative $\Delta (y_P)$ means that the net displacement of the cell is {\it down} the nutrient concentration gradient, which is opposite to what one expects for chemotaxis. To explain this behavior, we also measure $N_R(y_P)$ and $N_L(y_P)$ and plot their difference in Fig. \ref{S2_Fig}. This plot shows that $N_R(y_P) < N_L(y_P)$ for small $y_P$, {\sl i.e.} less number of rightward runs and more number of leftward runs are originated when $y_P$ is small. Note that a rightward run starting with a given $y_P$ must be preceded by a leftward run which terminates at the same $y_P$, and had originated at a lower $y_P$ value. This event becomes particularly unlikely when $y_P$ values are small, near the left-tail of the distribution $P_{tum}(y_P)$. Thus $N_R(y_P) < N_L(y_P)$ for small $y_P$, which makes $d_R(y_p) < d_L(y_p)$ and as a result, $\Delta (y_P) < 0$. As $y_P$ increases and comes out of the left-tail region of $P_{tum}(y_P)$, we find $N_R(y_P)$ gradually increases and overtakes $N_L(y_P)$, and $\Delta(y_P)$ becomes positive and keeps increasing with $y_P$. However, as one approaches the right-tail of the distribution $P_{tum}(y_P)$, both $N_R(y_P)$ and $N_L(y_P)$ tend to zero. Therefore, $\Delta(y_P)$ starts decreasing again for large $y_P$ and finally becomes zero. When we average $\Delta (y_P)$  over the distribution $P_{tum}(y_P)$ to calculate $\Delta$, small $y_P$ values give negative contribution and reduces $\Delta$. Note however, that negative $\Delta (y_P)$ values are near the left tail of $P_{tum}(y_P)$ and hence occur with low probability. Thus overall drift velocity still remains positive. 

The argument in the previous paragraph was presented for small $\lambda$ values. Even when $\lambda$ is larger, this argument can be applied (since our data in Fig. \ref{fig:chy}B show that for very small $y_P$ both rightward and leftward runs raise the $y_P$ level, but leftward runs do so by a larger magnitude). Thus even for larger $\lambda$ we have $N_R(y_P) < N_L(y_P)$ near the left-tail of the distribution $P_{tum}(y_P)$. However, for larger $\lambda$, $P_{tum}(y_P)$ becomes wider and its left-tail becomes longer (Fig. \ref{yp} inset). Thus negative values of $\Delta(y_P)$ are now observed for much smaller $y_P$ values. Averaging over such a curve yields a higher value of $\Delta$ than what was observed for small noise. This explains why $\Delta$ increases as noise increases. However, when noise becomes too large, the cell cannot distinguish between rightward and leftward runs. The change in activity or $y_P$ in a run is completely controlled by methylation level fluctuations now, and ligand concentration plays an insignificant role. Our data in Fig. \ref{fig:chy}C also shows that the two curves showing change in CheY-P level in a rightward and leftward run come closer as $\lambda$ increases. This again reduces the value of $\Delta$.

Above explanation and the accompanying data are presented for one dimensional motion of the cell with instantaneous tumbles. However, these arguments can be generalized for two dimensional case as well to explain our observation of optimal noise in that case.

\section{Discussion}
\label{sec:con}
In this paper, we have studied the effect of methylation noise on the chemotactic performance of a single {\sl E. coli} cell. We find that in the case of a nutrient environment that has spatial variation but no temporal variation, the chemotactic performance of the cell, measured in terms of localization and chemotactic drift velocity, shows a non-monotonic variation with the noise strength. There is an optimum noise strength where the best performance is observed. We explain this result from CheY-P level fluctuations for cell motion up and down the concentration gradient of the nutrient. We argue that for runs starting with a low CheY-P level, the cell shows a tendency to move down the nutrient concentration gradient which is detrimental to its chemotactic performance. The threshold value of CheY-P level below which this happens increases as the methylation noise strength is decreased. When we average over the CheY-P level statistics, for very low noise strength, the chemotactic performance is thus weaker. On the other hand, when the signaling noise is very large, the cell is unable to distinguish between runs up and down the gradient and its motion is totally controlled by stochastic methylation fluctuations. In this limit, the chemotactic performance is of course bad. Thus an intermediate noise level works best for the cell.

In the case when the nutrient environment has spatio-temporal variation, caused by diffusion and degradation of the nutrient in the medium, the chemotactic performance is best characterized by the first passage time and uptake. While first passage time shows a monotonic decrease with the methylation noise strength, the uptake may show a peak resulting from an interplay of time-scales associated with degradation and diffusion processes. We present the data for time-varying nutrient concentration in Appendix \ref{sec:time}.

In \cite{flores} it was shown that for a shallow ligand gradient, the chemotactic drift velocity shows a peak at a specific noise strength, while the localization remains constant at low noise level and decreases to zero as noise increases. The optimal noise level observed for drift was explained by using a simplified model where (a) the internal state of the signaling pathway is described just in terms of activity and both methylation level and CheY-P level are expressed as a function of activity, (b) tumblings are assumed to be instantaneous, (c) the sigmoidal nature of dependence of tumbling rate on activity was approximated by making the tumbling rate zero as the activity value falls below a threshold. Within this simplified model, it was shown that the drift motion results from the difference in the amount of time a right-mover and a left-mover spends in the small activity states. With increasing noise these small activity states are reached more often and hence drift velocity also increases. In comparison, our data in Fig. \ref{fig:yp} clearly show that the net displacement of the cell during a run can become negative when the run starts with a very low CheY-P level (or, equivalently, very low activity). Such runs make a negative contribution to the overall drift velocity and this important aspect should be taken into account while explaining increase of drift velocity with noise.

It should be possible to experimentally verify the existence of threshold CheY-P level that we predict from our model here. Monitoring the switching events from CCW to CW bias of flagellar motors, and keeping track of the cell position, one can obtain information about all the runs up and down the gradient. The CheY-P level at the tumbling event can be measured from the CW bias of the motors when the switching occurs. Using these data, it should be possible to determine $\Delta (y_P)$ experimentally and directly verify whether it becomes negative for low CW bias. Moreover, the methylation noise strength depends on the biochemical rate parameters and the total concentration levels of CheA, CheY and CheZ proteins \cite{frankel2014}. Thus even within an isogenic cell population, with identical pathway topology where biochemical rates are same, the protein concentrations can vary due to noisy gene expression. It will be interesting to experimentally measure localization or chemotactic drift velocity for different methylation noise strengths and verify the existence of an optimum noise level.

For the sake of simplicity, we have left out certain aspects of the signalling network from our model like adaptation of flagellar motors \cite{yuan12} or spatial organization of chemo-receptors \cite{emonet13}. Also, we have not considered any source of noise other than methylation level fluctuations \cite{victor2017}. Our understanding of the relationship between chemotactic performance and signaling noise within the  present simpler model will pave way for studying more complex models where above mentioned effects are included.

\section*{Acknowledgments}
SC acknowledges financial support from the Science and Engineering Research Board, India (Grant No. EMR/2016/001663). The computational facility used in this work was provided through the Thematic Unit of Excellence on Computational Materials Science, funded by Nanomission, Department of Science and Technology (India).

\appendix

\section{Search time for favorable region}
\label{sec:fpt}
\renewcommand{\theequation}{A-\arabic{equation}}
\setcounter{equation}{0}
In this section, we discuss how quickly a cell manages to find for the first time, the region with higher nutrient concentration. First passage time is the suitable measure in this case \cite{sdev}. Clearly, this is a response function measured away from the steady state. We measure the first passage time for different strengths of signaling noise in one and two dimensions.

\begin{figure}[!h]
\includegraphics[scale=0.6]{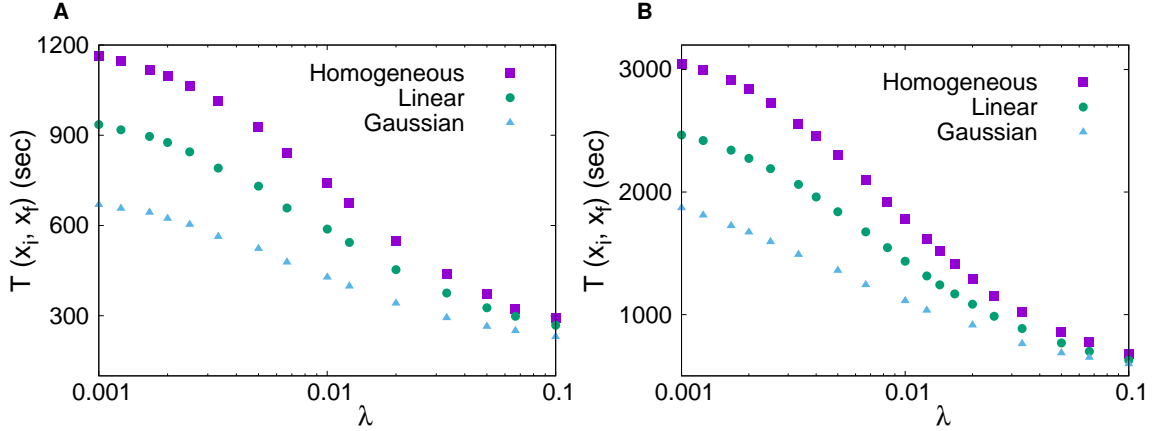}
\caption{{\bf Mean first passage time decreases as a function of noise strength.} 
{\bf (A)} shows the data in one dimension and {\bf (B)} shows the data in two dimensions. Square symbols correspond to a homogeneous concentration profile of the nutrient, $c(x)=c_0$,  circles correspond to a linear concentration form $c(x)=c_0(1+x/x_0)$; the triangles correspond to a Gaussian 
$c(x) = c_0(1+\frac{1}{\sqrt{2\pi\sigma^2}}\exp[-\frac{(x-\bar{x})^2}{2\sigma^2}])$ in both panels. We have used $c_0=200 \mu M$, $x_0=10^4 \mu m$, $\sigma=100 \mu m$, $\bar{x}=500 \mu m$. The mean first passage time $T(x_i, x_f)$ has been measured from an initial position $x_i=300 \mu m$ to a final target position at $x_f=490 \mu m$.}
\label{fvg} 
\end{figure}

In one dimension, we measure $T(x_i,x_f)$, defined as the time taken for a cell to reach a position $x_f$ for the first time, starting from an initial position $x_i$, where $c(x_f) > c(x_i)$. A small value of first passage time indicates an efficient search strategy. In Fig. \ref{fvg}A we plot the mean first passage time, which is averaged over different trajectories of the cell. We consider two different types of nutrient concentration profile: a linearly varying $c(x)=c_0(1+x/x_0)$ and a Gaussian $c(x)=c_0 \left (1+\dfrac{1}{\sqrt{2\pi\sigma^2}}\exp\left [ -\dfrac{(x-\bar{x})^2}{2\sigma^2} \right ] \right )$. For comparison, we also show the data for a homogeneous concentration profile $c(x)=c_0$ in the same plot. Our data in Fig. \ref{fvg}A show that in all cases the mean first passage time decreases as the noise strength increases. Also, for large noise values, the curves for the three different concentration profiles merge. In Fig. \ref{fvg}B we present data for two dimensions for homogeneous and linear $c(x)$. In this case, since $c(x)$ is independent of $y$-coordinate, the initial position of the cell has been taken anywhere on the line parallel to $y$-axis, with $x$-coordinate $x_i=300 \mu m$. Similarly, the target position is another parallel line with $x_f=490 \mu m$. We find similar behavior as in the one dimensional case here, although the values of the first passage time are larger in two dimensions. 

As follows from our data in Fig. \ref{power}, for large noise, the long runs become more probable, which clearly help the cell to explore the medium quickly. As a result, it becomes possible for the cell to cover the distance to the target in a relatively small number of long runs. On the other hand, when the noise is small, run length distribution falls off exponentially and probability to observe long runs is negligible. In this case, the cell tumbles rather often and before it hits the target, a large number of tumbles and hence directional changes have been executed. In this limit, the mean first passage time is longer and can actually be calculated analytically in one dimension \cite{sdev}.

From our data in Fig. \ref{fvg} we therefore conclude that as noise increases, the search process becomes quicker. However, a very large noise brings about wild fluctuations of the protein levels and that is bound to affect the chemotactic performance in an adverse way. Long runs may be good for the cell to explore the whole environment quickly, but a good chemotactic performance in the long time limit actually demands that the cell is able to distinguish between runs up and down the concentration gradient. A strong fluctuation in the methylation level causes that distinction to become blurred since the change in activity in Eq. \ref{eq:act} is now dominantly controlled by the methylation level changes, rather than local change in the ligand concentration. Indeed our data in Sec. \ref{sec:space} show that the steady state chemotactic performance becomes poor when the noise is very large.

\section{Effect of signaling noise with time-varying nutrient concentration}
\label{sec:time}
\renewcommand{\theequation}{D-\arabic{equation}}
\setcounter{equation}{0}
In many physical situations the nutrient environment experienced by the cell changes with time. The chemo-attractant may undergo diffusion in the medium which reduces its concentration gradient as time goes on, or it can degrade into some other chemical which can not be sensed by the chemo-receptors of the cell and during this degradation process the overall concentration level of the chemo-attractant drops with time. In this situation, it is immaterial whether the cell can position itself preferentially in the nutrient-rich region in the long-time limit, because the nutrient may not last till that long time. Rather, the trajectory of the cell in the medium in the relatively short time regime, while the nutrient concentration gradient is still present, is more crucial. In an harsh environment, when the nutrient is depleting fast from the medium, the cell needs to find the nutrient-rich spots quickly and also its trajectory in the medium should be such that it encounters a large number of nutrient molecules.

In this section, we consider a situation where a certain amount of chemo-attractant or nutrient is injected in the medium at a spatial location $\bar{x}$, following which the chemo-attractant undergoes diffusion and degradation in the medium. Then the  concentration profile of the chemo-attractant has the shape of a Gaussian. The  width of this Gaussian function keeps increasing with time. The background concentration level of the nutrient keeps falling with time. After a while, when the width of the Gaussian reaches a value $\sigma_0$, a bacterial cell is introduced in the medium at a position $x_0$. The nutrient concentration experienced by the cell at a position $x$ at time $t$ after its introduction is then given by 
\be
c(x,t)= c_0 e^{-t/\tau_d} \left [ 1+ \frac{\exp { \left ( -\dfrac{(x-\overline{x})^2}{\sigma_0^2 + 4{\mathcal D}t } \right )}}{ \sqrt{2 \pi ( \sigma_0^2 + 4{\mathcal D}t)}} \right ], \label{eq:cxt}
\ee  
where ${\mathcal D}$ is the nutrient diffusivity, $\tau_d$ is the time-scale of nutrient degradation. The chemotactic performance in this case is measured by (a) the  first passage time of the cell measured at a region close to the peak of the Gaussian where the nutrient concentration is highest, and (b) uptake, defined as the mean amount of nutrient encountered by the cell along its trajectory up to a large enough observation time \cite{celani2010} . This is measured by the quantity ${\mathcal U}=\int_0 ^{t_{obs}} dt \int_0^L dx c(x,t) {\mathcal P}_{\lambda} (x,t) $. Note that due to degradation of the nutrient, the integrand vanishes for $t \gg \tau_d$ and the uptake saturates to a finite value, even as $t_{obs}$ is increased. We examine the dependence of first passage time and uptake on the signaling noise. We limit our studies to one dimension.

\subsection{Decaying nutrient profile}
\begin{figure}[!h]
\includegraphics[scale=1.2]{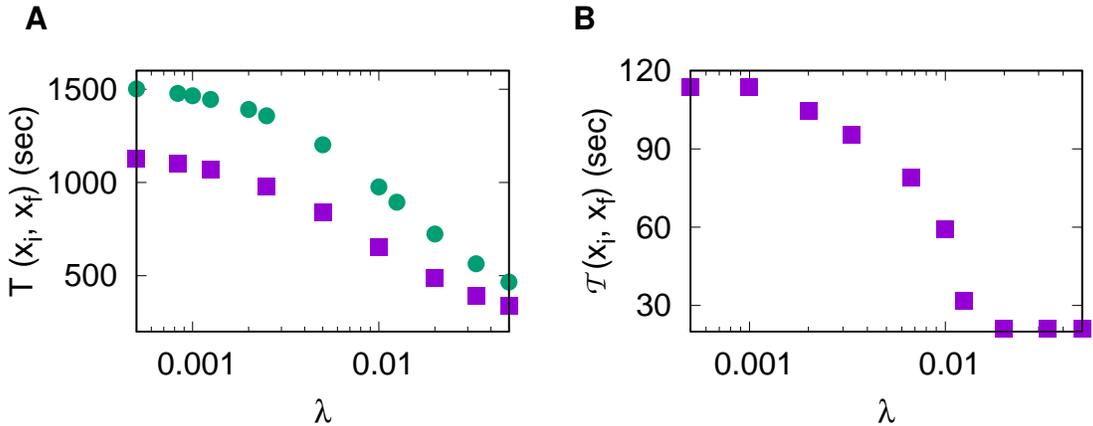}
\caption{{\bf First passage time vs noise strength for degrading nutrient profile.} {\bf (A)}: Mean first passage time $T(x_0,x_f)$ decreases with $\lambda$. The mean first passage time has higher values than that in Fig. \ref{fvg} where no degradation is considered. {\bf (B)}: The typical first passage time ${\mathcal T}(x_i,x_f)$ has much smaller value than the mean, but also decreases with $\lambda$. The circles show the data for $c(x,t) = c_0 e^{-t/\tau_d}$ and squares are for $c(x,t)$ given by Eq. \ref{eq:cxt} with ${\mathcal D} =0$. These data are for the one dimensional case and we have used $\tau_d=500 sec$, $\sigma_0 = 100 \mu m$. Other parameters are as in Fig. \ref{fvg}A.}
\label{fpt_decay}
\end{figure}

First we consider the limit when ${\cal D}$ is very small. In this case, the cell experiences a Gaussian concentration profile with almost fixed width $\sigma_0$, and an exponential decay of the overall concentration level. Starting from an initial position $x_i$ we measure the time taken by the cell to reach the peak region around $\bar{x}$ for the first time. Clearly, this first passage time is a stochastic variable and for different trajectories of the cell it takes different values. The probability distribution of the first passage time generally has a long tail which makes the mean much larger than the most probable or typical value \cite{redner,metzler}.  In Fig. \ref{fpt_decay}A we plot the mean first passage time as a function of methylation noise strength $\lambda$. We find that the mean first passage time decreases with  $\lambda$. This qualitative behavior is similar to our observation in Fig. \ref{fvg}A when there was no degradation of the nutrient. However, a quantitative comparison between Figs. \ref{fvg}A and \ref{fpt_decay}A  shows that when the nutrient degrades, the first passage time takes higher value for the same noise strength. This effect is visible even when the nutrient concentration is homogeneous. A simple way to understand why degradation of nutrient makes the search process slower is that when nutrient degrades, even when the cell is moving in a homogeneous medium, it experiences a decreasing concentration along its trajectory, which makes it tumble more. This makes the average run durations shorter and hence the mean first passage time longer \cite{sdev}. Note that for small $\lambda$, the actual value of the mean first passage time is much longer than the degradation time-scale $\tau_d$. In this case, the typical first passage time, which is much smaller than the mean first passage time, is perhaps a more suitable measure of chemotactic performance. In Fig. \ref{fpt_decay}B we plot the typical first passage time as a function of $\lambda$ and find that just like the mean, the typical value also shows similar qualitative dependence on noise. 

\begin{figure} [!h]
\includegraphics[scale=1]{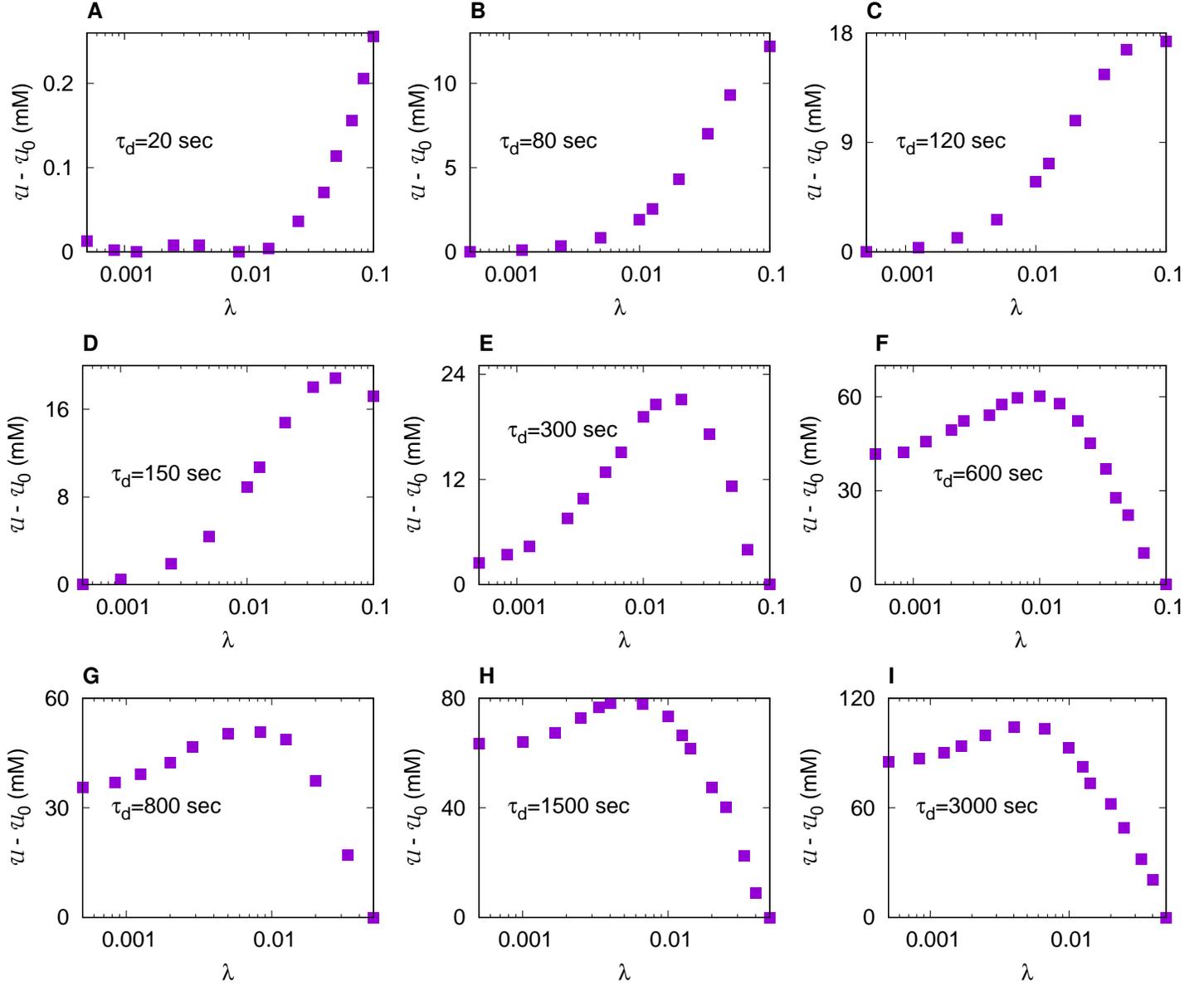}
\caption{{\bf Change in uptake as a function of noise strength for decaying Gaussian concentration profile}: {\bf (A-C):} For fast degradation of the nutrient, uptake increases with $\lambda$, since maximum contribution to $\mathcal U$ in this case comes from those trajectories with small first passage times. {\bf (D-G):} For a slower degradation of the nutrient, uptake starts decreasing for very large noise, and a peak is observed. Reaching the peak in the shortest possible time is not the single most important criterion any more. When the nutrient lasts for some time, those trajectories where the cell takes longer to reach the peak and then spends longer time in the peak region, contribute more towards uptake. {\bf (H-I):} For a very slow degradation, the peak of uptake moves closer to the optimal noise $\lambda^\ast$, observed in Fig. \ref{loc}B. The values of $\mathcal U_0$ used in panels {\bf (A-I}) are $406$, $1628 $, $2449$, $3064$, $5956$, $10023$, $11782$, $15068$, $17551$ $mM$, respectively. These data are for one dimensional system and we have used $t_{obs} = 1000 sec$. Other simulation parameters are as in Fig. \ref{fvg}A.}
\label{uvgde}
\end{figure}

The uptake $\mathcal U$ measures the amount of nutrient intercepted by the cell along its trajectory up to a certain observation time $t_{obs}$. We find that the uptake may increase or decrease monotonically with noise, or may even show a peak, depending on the degradation time-scale $\tau_d$. For small values of $\tau_d$'s we find that uptake increases with noise (Fig. \ref{uvgde}A-C). In this case, when the nutrient degrades very fast, the maximum contribution from uptake comes from those trajectories with very long runs, which enable the cell to reach the peak of the Gaussian before the nutrient had degraded significantly. As $\lambda$ increases, the probability of such long runs increases and hence uptake also increases. On the other hand, when $\tau_d$ is large, then the degradation happens slowly and within the large but finite observation time $t_{obs}$, not much degradation has taken place. In this limit we expect to recover the results for the time-independent nutrient environment. Indeed our data in Fig. \ref{uvgde}D-I show that uptake develops peak at  particular $\lambda$ values and as $\tau_d$ increases the peak position approaches the optimal $\lambda ^ \ast$ observed in Fig. \ref{loc}B.

\subsection{Nutrient profile with decay and diffusion}
In the case when the nutrient diffusivity $\mathcal D$ is not so small, the nutrient diffusion cannot be neglected over the time-scale of cell movement, and the cell experiences the full time-dependent nutrient profile with decay and diffusion, given in Eq. \ref{eq:cxt}. For a given value of $\mathcal D$ and $\tau_d$, we find same qualitative behavior for the first passage time, as found in the previous subsection for negligible $\mathcal D$ (data not shown).

\begin{figure}[!h]
\includegraphics[scale=1]{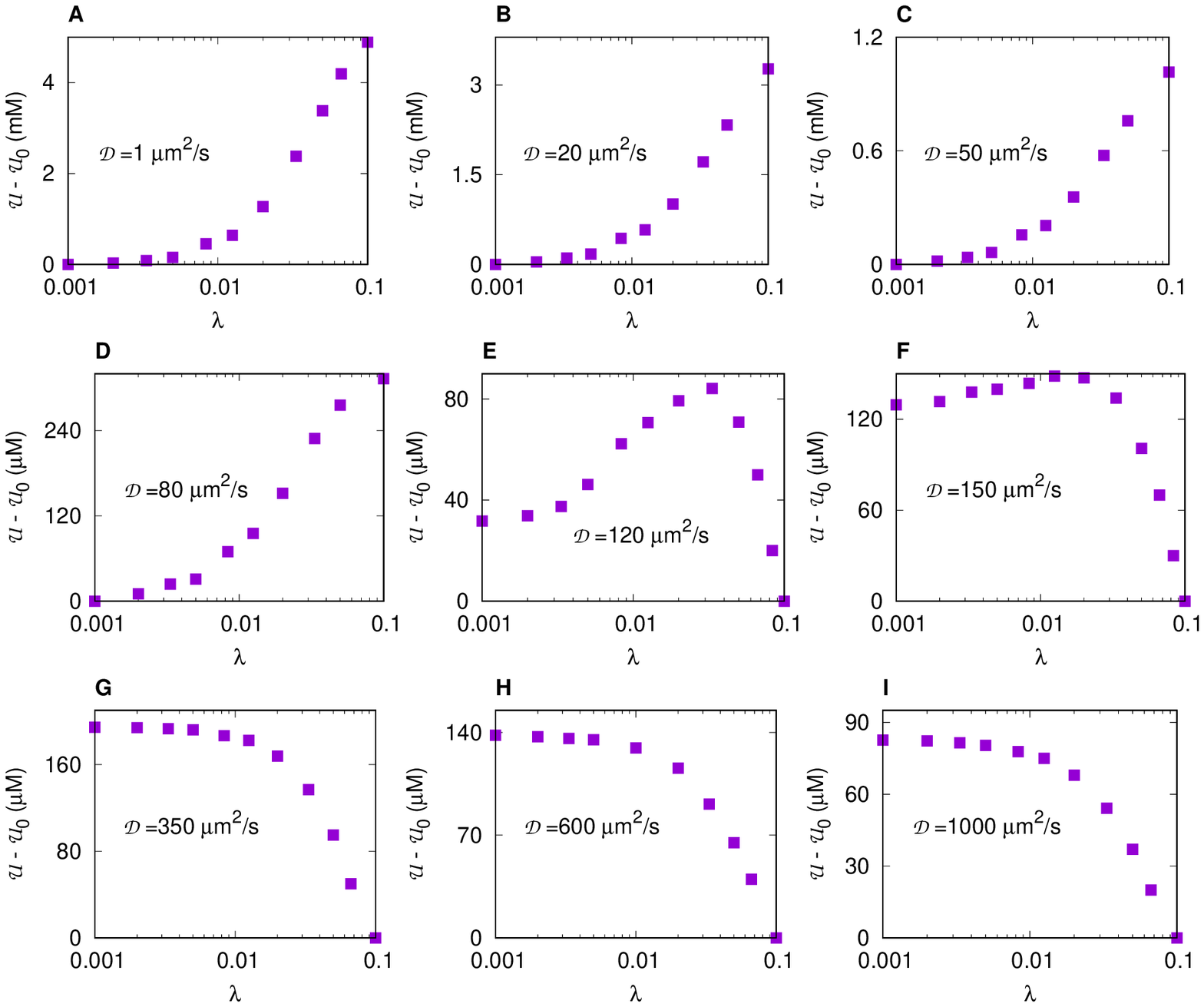}
\caption{{\bf Change in uptake as a function of noise strength for nutrient profile with diffusion and degradation.} {\bf (A-D):} For slow diffusion of the nutrient, uptake increases with noise for our choice of $\tau_d$. In this case, uptake is governed by those trajectories where cell executes long runs. {\bf (E-F):} For a faster diffusion of nutrient, uptake shows a peak as a function of $\lambda$. But the variation is much weaker. {\bf (G-I):} For very fast diffusion of nutrient, uptake decreases with noise. The values of $\mathcal U_0$ used are: $1730$, $1736$, $1738$ $mM$ in panels {\bf A,B,I}, respectively, $1740 mM$ in panels {\bf C,H}, $1741 mM$ in {\bf G}, and $1742 mM$ in {\bf D,E,F}. We have used one dimensional system and the nutrient concentration in Eq. \ref{eq:cxt} with $\tau_d=100 sec$, $\sigma=10 \mu m$, $t_{obs} = 200 sec$ here. Other simulation parameters are same as Fig. \ref{fvg}A.}
 \label{uvgdide} \end{figure}

However, uptake shows interesting difference depending on the choice of $\mathcal D$. For a given value of $\tau_d$, when $\mathcal D$ is very small, we recover the results of the previous subsection. In this regime, the behavior of uptake is controlled by $\tau_d$. For our choice of $\tau_d=100s$, we find that (see Figs. \ref{uvgdide}A-D) uptake increases with noise when $\mathcal D$ is small, in agreement with Fig. \ref{uvgde}A-C. On the other hand, when $\mathcal D$ is very large, then the nutrient diffuses very fast and soon the concentration gradient in the medium disappears and the dependence of uptake on cell trajectory ceases to exist. The variation of uptake with $\lambda$ is much weaker in this case. The short time dynamics of the cell in this case decides the uptake variation. The more time the cell is able to spend close to the peak of the Gaussian profile before the profile flattens or nutrient degrades, larger will be its uptake. When $\lambda$ is large, the cell executes long runs and has shorter first passage time to the peak region, which reduces its residence time in the region that lies in between its initial position and the peak. So uptake is small for large $\lambda$. But when $\lambda$ is small, the first passage time is larger and the cell spends most of its short time trajectory trying to climb up the concentration gradient, reaching the peak of the Gaussian profile. This increases the uptake. Thus for very large $\mathcal D$ uptake decreases with $\lambda$, as shown in  Figs. \ref{uvgdide}G-I. Therefore, for  intermediate $\mathcal D$'s uptake must show a peak with $\lambda$ [Figs. \ref{uvgdide}E and \ref{uvgdide}F].  



\end{document}